# Damping Analysis of Subsynchronous Oscillations at the Sending-End of Practical Grid-Forming MTDC Power System for Isolated Renewable Energy

Qiang Fu, *Member IEEE*, Siqi Bu, *Senior Member IEEE*, and Xiao Chen

***Abstract*—Subsynchronous oscillations (SSOs) have occurred at the sending-end of Zhangbei grid-forming MTDC power system (SE-GFPS). This paper gives an alert for the wide application of grid-forming (GFM) control by presenting a detailed report on such SSOs and the practical GFM configuration. Different from traditional studies that consider only parts of GFM control loops for simplification, power synchronous control (PSC), *d*- and *q*-axis AC voltage control (AVC), and inner current control (ICC) loops are comprehensively involved in this study. A self- and coupling-damping method is proposed to quantify the impact of both the inherent dynamics of different GFM control loops and the external dynamic coupling between the GFM control and the remaining SE-GFPS on the SSOs. It determines whether the major causes of SSOs are attributed to inherent GFM dynamics or external dynamic couplings. Based on the damping sensitivity analysis, the major impact factors of the SSOs are identified. Self-damping can be improved more by *q*-axis than by *d*-axis AVC parameters, and negative coupling damping can be reduced by PSC parameters. Finally, SSO mitigation strategies are proposed, and an SE-GFPS mirroring the real-world Zhangbei project is established on the electromagnetic transient platform in Simulink, validating the accuracy of our conclusions.**

***Index Terms*—Grid-forming control, renewable energy, voltage source converter (VSC), subsynchronous oscillations, small signal stability, Zhangbei MTDC project**

## I. INTRODUCTION

To support the deep decarbonization of the global energy system, the wind and solar power expand rapidly such that the share of renewable energy in global primary energy increases from around 10 % in 2019 to between 35 % and 65 % by 2050 [1]. Motivated by this, the share of renewable energy is predicted to increase significantly.

However, a large number of wind farms are far away from load centers, such as offshore wind farms and the north-west wind farms in China [2]-[3]. To address these issues, high voltage alternating current (HVAC) and high voltage direct current (HVDC) transmission techniques have been proposed for wind power delivery [4]-[5]. It has been demonstrated that using the HVDC technique has many advantages than using the HVAC technique under the condition of long-distance power transmission, such as higher controllability and lower cost [6]. Therefore, the HVDC transmission system is more attractive for power delivery. The maximum transmitted power of the line commutated converter (LCC) is larger than that of the voltage source converter (VSC), while the LCC has more limitations, such as it cannot relate to weak power systems and needs much reactive power to support AC voltage [7]. Therefore, the VSC is a better choice for wind farms in isolated regions with weak AC voltage support ability. With these advantages, VSC-based HVDC (VSC-HVDC) and VSC-based multiple terminal HVDC (VSC-MTDC) projects have been established worldwide, and this study is motivated by one of them.

Qiang Fu and Siqi Bu are with the Department of Electrical and Electronic Engineering at The Hong Kong Polytechnic University. Xiao Chen is with the North China Branch of State Grid Corporation of China and is responsible for data provision. This work is supported by the National Natural Science Foundation of China (Grant No. 52407129), and the Supported by Sichuan Science and Technology Program (Grant No. 2025YFHZ0234).
Corresponding author: Siqi Bu, Email: siqi.bu@polyu.edu.hk.

### *A. Unique Features of the Power System at the Sending-end of Zhangbei Grid-Forming MTDC Power System (SE-GFPS)*

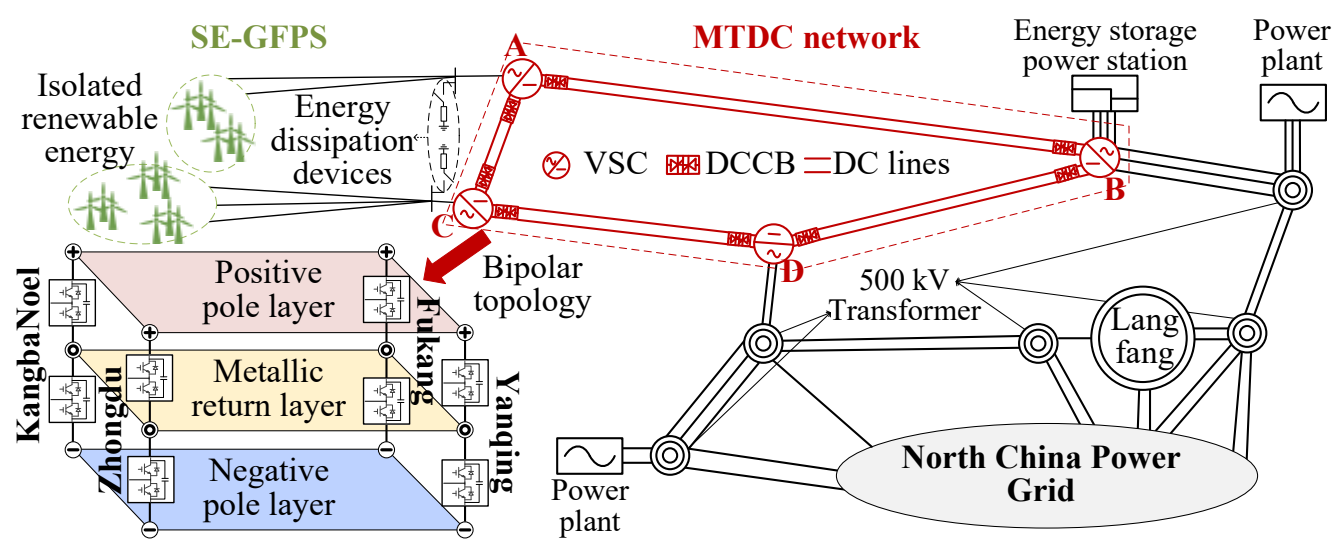

Fig. 1. Configuration of the Zhangbei MTDC project (VSC stations are labeled as: A (KangbaNoel), B (Fukang), C (Zhongdu), and D (Yanqing)).

As illustrated in Fig. 1, the Zhangbei MTDC project is the first meshed VSC-based MTDC transmission system in the world. The DC voltage is ± 500 kV, and the rated power is 9000 MW, which remains the largest MTDC project in the world as of 2025. The DC circuit breakers (DCCBs), represented by red blocks, are installed at both terminals of each DC transmission line. A major feature of the Zhangbei MTDC project is that the VSCs A and C at the sending-end adopt GFM control. The GFM-VSCs actively control AC voltage and frequency of the SE-GFPS as constant under the condition of "No SGs". For clarification, the differences between the SE-GFPS focused on in this paper and traditional sending-end AC power systems (SEPSs) dominated by SGs are presented in Table I.

TABLE I
Differences between SE-GFPS in This Study and Traditional SEPSs.

| Refs. | Control types | | Applied scenarios | | | Features of SEPSs | |
|---|---|---|---|---|---|---|---|
| | GFM | GFL | ES | RE | HVDC | No SGs | With SGs |
| [8] | | ✓ | ✓ | | | | ✓ |
| [9] | | ✓ | | ✓ | | | ✓ |
| [10] | | ✓ | | | ✓ | | ✓ |
| [11] | ✓ | | ✓ | | | | ✓ |
| [12] | ✓ | | | ✓ | | | ✓ |
| [13] | ✓ | | | ✓ | | | ✓ |
| **This study** | ✓ | | | | ✓ | ✓ | |

Notes: GFL: Grid-Following, ES: Energy Storage, RE: Renewable Energy.
No SGs: This indicates that the SEPS does not include synchronous generators (SGs), the voltage and frequency are controlled by GFM-VSC.
With SGs: This indicates that the frequency and voltage of the power system have been well controlled by SGs, even without GFM-VSC.

It has been well known that the dynamics of GFM and GFL controls differ [15]. Additionally, the SE-GFPS in this study comprises entirely RE, where the frequency and voltage are fully controlled by a GFM-VSC. However, in traditional SEPSs dominated by SGs, GFM-VSC is used as a coordination in frequency and voltage adjustments. Therefore, the dynamics of SE-GFPS are significantly different from traditional systems, both in VSC control and system-level dynamic characteristics.

### *B. Simplifications in the Dynamics of GFM Control Loops*

There are various specific configurations of GFM control [17], resulting in different dynamics. To simplify the stability analysis, parts of GFM control loops are often neglected. We categorize these simplifications into three types, as summarized in Table II and illustrated in Fig. 2.

TABLE II
Scopes of Retained Dynamics in GFM Control Loops

| Scope of retained dynamics | Small signal stability | | Case studies |
|---|---|---|---|
| | EMOs | SSOs | |
| Dynamics of PSC are retained, while those of AVC and ICC control loops are ignored (Scope 1) | [19] | | HS |
| | [20] | | HS |
| | [21] | | RS |
| | [23] | [22] | HS |
| Dynamics of PSC and AVC are retained, while those of ICC control loops are ignored (Scope 2) | | [24] | HS |
| | [25] | | HS |
| | | [26] | HS |
| | | [27] | HS |
| | [28] | | HS |
| Dynamics of all control loops are retained (Scope 3) | | [29] | HS |
| | | [30] | HS |
| | | **This study** | **RS** |

Notes: HS: The investigations are carried out based on hypothetical systems.
RS: The investigations are conducted using real-world systems.

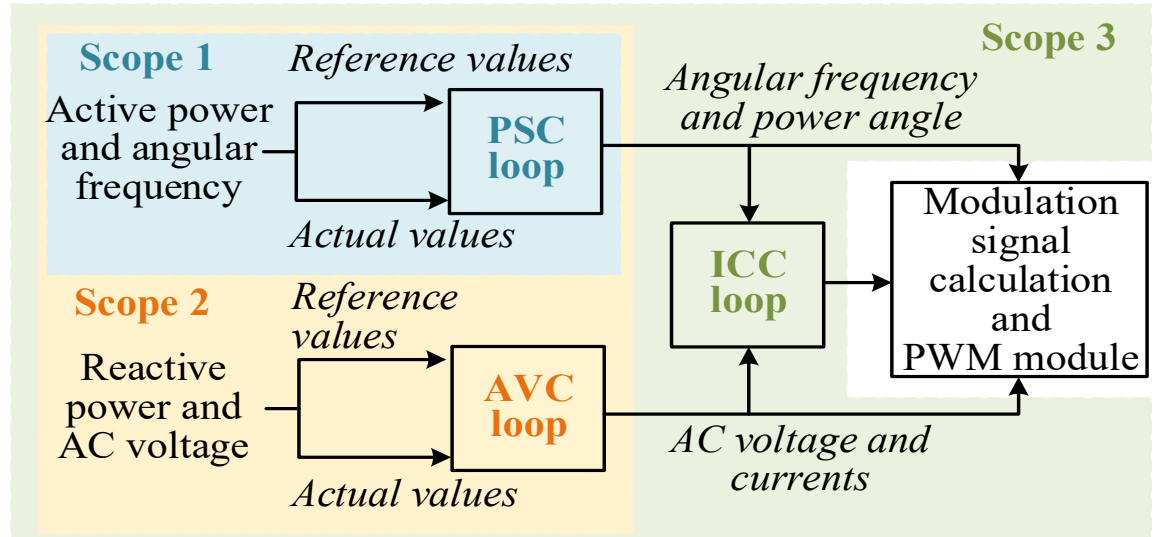

Fig. 2. Illustration of the scopes of retained dynamics in GFM control loops.

Scope 1: Only the power synchronous control (PSC) loop is considered, such that the highest order of the GFM control is no more than second-order in normal, such as virtual synchronous generator (VSG) control and frequency droop control [18]. The dynamic equations of VSG control are similar to those of SGs; thus, electromechanical oscillations (EMOs) may occur. In [19] and [20], the occurrence of EMOs caused by VSG control was confirmed using the Argentinian and New England networks as study cases, respectively. However, these findings are limited to theoretical analysis, and there have been no reported incidents of such occurrences in practical projects until now. Instead, authors in [21] shared operational insights from the first grid-connected wind farm utilizing GFM control in the UK. Their findings illustrated that the VSC with GFM control can enhance the inertia of AC power systems. Considering that the parameters of PSC loop can be adjusted as needed, SSOs can also be caused by GFM control [22].

Scope 2: Both PSC and AC voltage control (AVC) loops are considered. For general GFM control, the dynamics of PSC and AVC loops are coupled [23]. Therefore, considering the coupling dynamics makes the results more reliable. In [24], a control interaction model considering both PSC and AVC loops was established, and critical parameters that dominate dynamic interactions between the ES with GFM control and wind power plants were clarified. In [25], the impact of AVC loop on frequency stability is considered in dynamic equations of PSC loop, simplifying the transfer function from multiple-input and multiple-output (MIMO) into single-input and single-output (SISO). In [26], the voltage and frequency stability were investigated by equating the dynamics of PSC and VAC loops as a SISO complex impedance, making the impedance-based method applicable. The dynamic coupling issue is severe in stiff grids, pushing the power system including with GFM control to become unstable. To address this issue, a dynamic power-decoupling method was proposed in [27]. Furthermore, strong interaction between any two subsystems is another major risky factor, such as in [28], the modal resonance between multiple GFM-controlled converters was demonstrated, and a method of coordinately adjusting parameters was proposed.

Scope 3: All control loops, including PSC, AVC, and inner current control (ICC) loops, are considered. In [29], a full-order model of the power system using GFM control with all control loops considered was established, and the dynamics were analyzed using modal analysis methods. However, they only provided numerical calculation results without mechanistic explanations, making it difficult to clarify the general risky conditions of GFM control. In [30], simulations of Scope 1 and Scope 2 were also illustrated, and the impact of ICC on the stability of GFM converters was analyzed. It was concluded that improper selection of models can result in errors because no single reduced model is suitable for all scenarios. Therefore, considering the complete structure of GFM control instead of ignoring them is more suitable for practical instability analysis.

### *C. Aims and Contributions*

This study is motivated by the practical implementation of GFM control for grid connection and aims to raise awareness among researchers and engineers regarding instability risks caused by GFM control in grid-level applications. The major contributions are summarized as follows:

1) This paper presents practical insights into an SE-GFPS within confidentiality constraints. The SE-GFPS is primarily powered by RE and is dominated by a GFM-VSC, as opposed to traditional SGs. In this case, a practical configuration of GFM control is studied, including PSC, AVC, and ICC loops in *d*-*q* coordinates. It is more practical than traditional methods that only consider PSC and simplified AVC loops, and provides a more reliable basis for grid-level stability investigations on GFM control compared to theoretical cases.

2) We propose a self- and coupling-damping analysis method that considers the inherent dynamics of GFM control and the dynamic coupling between the GFM-VSC and the remaining SE-GFPS. This method proposes self-damping and coupling-damping to quantify how parameters of inherent GFM control and the external transmission line independently affect the SSO damping. This highlights that coupling-damping is negative and causes the instability, even when the inherent GFM control parameters have been well designed.

3) We conduct a sensitivity analysis of self- and coupling-damping to identify the main contributors to SSOs. The results indicate that the SSOs in the SE-GFPS are primarily affected by the parameters of $q$-axis AVC loop and the power amplitude of the SE-GFPS. Accordingly, an additional damping control for the GFM-VSC and an adaptive control for GFM control parameter tuning are proposed and examined. It is theoretically verified that the underlying mechanism through which these controls enhance stability is that they equivalently improve the parameters of the AVC loop.

The remaining sections of this manuscript are organized as follows. In Section II, we report the practical examinations and determine the focused power system. Then, Section III briefly presents the mathematical model of the SE-GFPS. In Section IV, we reveal the mechanism of the SSOs, identify the primary reasons, and propose mitigation methods based on the proposed self- and coupling damping analysis method. In Section V, we recreate an SE-GFPS similar to the Zhangbei project based on the Simulink electromagnetic transient (EMT) platform, which reproduces the SSOs and validates accuracy of conclusions. Finally, Section VI summarizes the conclusions of this study.

## II. Report of Practical SSOs and Clarification of the Focused Power System for Theoretical Analysis

### *A. Report on the SSOs*

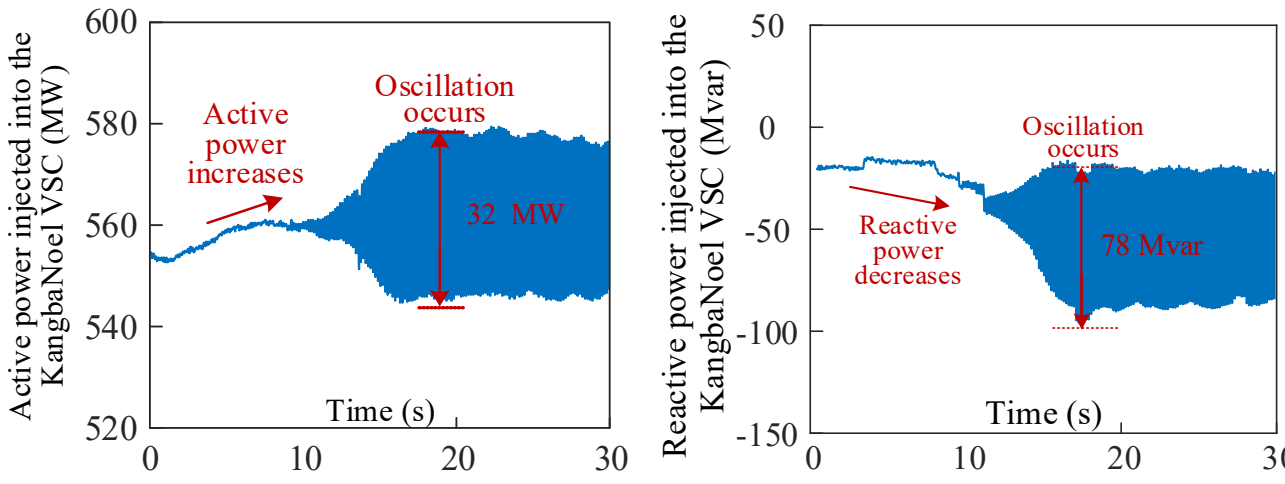

Fig. 3. Detected oscillation curves.

Converters in Zhangbei MTDC project can independently operate in either monopolar or bipolar mode, thereby enhancing the flexibility and reliability. In 2022, the Zhangbei MTDC project operated in the monopolar mode. During this period, an oscillation was detected in the SE-GFPS (KangbaNoel) when the output power exceeded 550 MW, reaching approximately 70 % of the rated monopolar capacity (750 MW). The detected oscillation curves are depicted in Fig. 3, where the oscillation frequency of active and reactive power curves is 6 Hz. Thus, this instability is identified as a subsynchronous oscillation (SSO) in this study. In Fig. 3, the SE-GFPS did not experience a large disturbance, and oscillation occurred around the steady-state operating point. Therefore, the stability issue focused on in this paper is the small signal stability of the SE-GFPS.

Importantly, this represents the first occurrence of an SSO related to a practical SE-GFPS. Therefore, it is imperative to precisely identify the cause and offer insights to comprehend its formation, rather than relying on numerical calculations, which is the primary motivation of this study.

### *B. Examinations on Causes of the SSOs in Practice*

Some examinations have been conducted in practice to identify the region of these instability risk factors, which can help narrow down the investigation scope of the SSO causes, thereby reducing the analysis complexity. Detailed information is summarized in Fig. 4 and reported as follows:

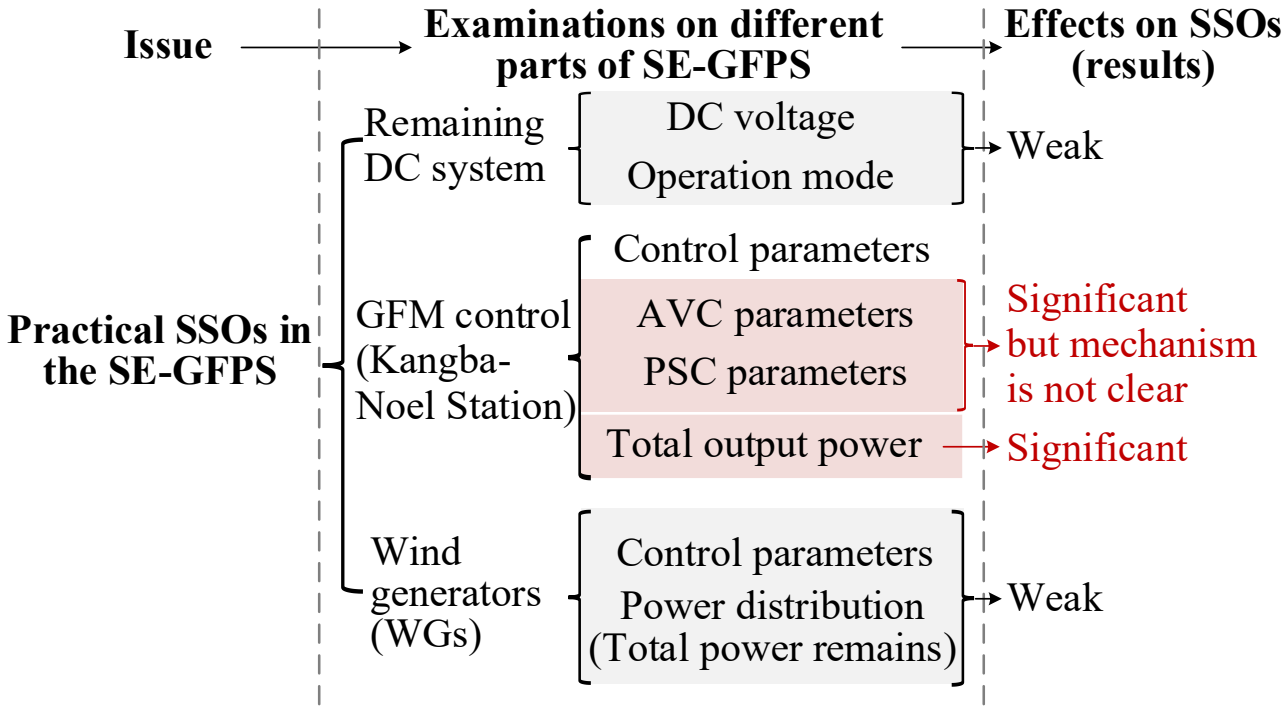

Fig. 4. Summary of the examined results.

1) Examination to determine whether the SSOs are caused by the DC network and then propagate to KangbaNoel Station.

To assess whether the oscillation is related to the remaining DC network, the rated DC voltage was changed from 500 kV to ± 500 kV (the operation mode was switched from monopolar to bipolar ). However, the same oscillation occurred when the active power reached ~70 % of the capacity of KangbaNoel Station. This indicates that increasing DC voltage can transmit more active power because the rated power of KangbaNoel Station is increased, but cannot eliminate the oscillation. The active power output from KangbaNoel Station is limited to values less than or equal to 0.7 per unit (p.u.), regardless of DC-side operation mode. Additionally, the oscillation is significant at the AC-side of KangbaNoel instead of Fukang Station, which implies that the oscillation is not caused by the DC network (if so, there would be oscillations in Fukang Station also).

2) Examination to determine whether the SSOs are caused by wind generators and then propagate to KangbaNoel Station.

Considering there have been many instances of SSOs caused by wind generators [31], some examinations were conducted to analyze whether this instability is caused by wind generators as well. First, the active power output from different wind farms were adjusted to examine whether changing the operating states of different wind farms can eliminate the oscillation. However, the oscillation still reappeared at the same frequency when the total output power reached ~0.7 p.u., regardless of the operating state of different wind farms. Second, the impedance curves of wind generators were measured (given in [16]). It can be seen that changing the operating states of wind generators induced weak effects on their dynamics at ~6 Hz. This explains why altering the operating state of each wind generator weakly affects the oscillation. Therefore, the oscillation was not fully generated by the wind generators.

3) Examination to determine whether the SSOs are caused by GFM control of KangbaNoel Station.

The control parameters of KangbaNoel Station were tuned to observe whether the SSOs disappeared. However, there was no idea about which control parameter should be tuned, because this incident occurred for the first time. To obtain results to identify some high-risk parameters, control parameters were examined one by one using the trial-and-error method. These examinations were based on the full EMT simulation platform, which is the same as the practical system. However, the simulation system is so complex that the examination requires several days. It is not sustainable to examine parameters serially (one-by-one) without any guidance. Therefore, it can only provide inspiration for possible control parameters instead of identifying accurate relations between the control parameters and instability.

Based on the above examinations, the SSOs are identified as originating within the SE-GFPS, highly related to GFM-VSC, and weakly coupled with the remaining DC power system. The high-risk factors are highlighted in red in Fig. 4. Therefore, a clear understanding of why the SSOs occur and how to properly design the GFM control parameters when connecting isolated RE sources is an essential research topic.

### *C. Focused Power System for Theoretical Analysis*

The examinations in practice are meaningful for identifying the key components and factors contributing to the SSOs. They indicate that the GFM-VSC plays a critical role and should not be simplified. Conversely, other less critical components can be simplified without significantly affecting the analysis results. The relationships between the actual project and this study are detailed in Table III.

The configuration of power system under focus is illustrated in Fig. 5 in blue. VSCs 1 and 2 adopt GFM control, which regulates the exchanged power between the DC and AC power systems. VSC-4 controls the DC voltage in the usual manner [32]. The AC-side of VSCs 3 and 4 is represented by an infinite bus for simplicity, as it is not the focus of this study. $X_f$ and $X_L$ denote the reactance of converter filter and transmission line, respectively. $V_{dc1}$ represents the DC voltage of VSC-1. $P_T$ and $Q_T$ represent the active and reactive powers output from VSC-1, respectively. $V_p$, $V_c$, and $V_0$ correspond to the AC voltages at the point of common coupling (PCC), the terminal of VSC-1, and the terminal of the AC power system, respectively. The subscripts $d$ and $q$ refer to the $d$-axis and $q$-axis components of variables, respectively. The subscripts $x$ and $y$ denote the $x$-axis and $y$-axis components of variables, respectively. The $d$-$q$ coordinates are established on the control loops of the VSC-1, and the $x$-$y$ coordinates are established on the AC power system. Both are two-phase rotating coordinate systems with a frequency of $\omega_0$ = 100 π rad/s. The power flow direction is indicated by the black arrow, and the positive direction of variables is aligned with the power injection direction from the VSC-1 into the AC power system. Consequently, the values of power and currents are negative for the sending-end power system in this study.

TABLE III
Relationships between The Actual project and This Study

| Items | | Actuality | This Study |
|---|---|---|---|
| Stability type | | SSOs at ~ 560 MW | Small signal stability |
| **Explanations** | | As the SSOs occur, the steady-state of power system is around 560 MW without any significant faults. | |
| Control type | VSC-1,2 | V/F control (GFM) | V/F control (GFM) |
| | VSC-3 | Power control (GFL) | Power control (GFL) |
| | VSC-4 | DC voltage control (GFL) | DC voltage control (GFL) |
| **Explanations** | | Dynamics of GFM-VSC is major, and thus all control loops are retained. | |
| Converter type | | Modular multilevel converter (MMC) | Average model-based converter and MMC |
| **Explanations** | | The oscillation frequency is 6 Hz, the average model is accurate enough for theoretical studies. The simulation uses the MMC for precise verification. | |
| Sending-end power system | | Various renewable energies | Controlled power and voltage source |
| **Explanations** | | In examinations, GFM-VSC is identified as the major cause, the SSOs are weakly related to the power distribution of wind generators, thus they are represented by a controlled power and voltage source. | |
| Receiving-end power system | | North China Power Grid | Infinite bus |
| **Explanations** | | It does not participate in the oscillation. | |

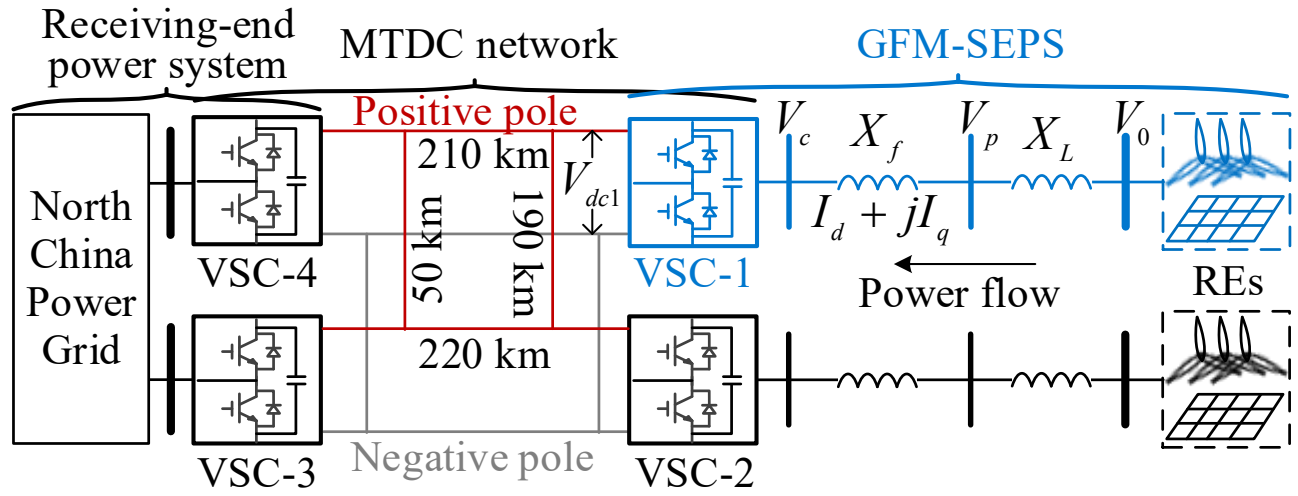

Fig. 5. Configuration of the focused power system.

## III. LINEARIZED MODEL OF THE SE-GFPS

In Fig. 6, the configuration of the GFM control that has been adopted in KangbaNoel Station is illustrated. The AVC, ICC, and PSC loops are highlighted in blue, grey, and red, respectively. The superscript $^{ref}$ means the reference values of control systems. Δ means a small increment in variable. $f_0$ is the fundamental frequency of AC power system, i.e., 50 Hz. $\theta$ is the voltage angle of $V_p$. The capital letter $K$ is used to represent various control parameters.

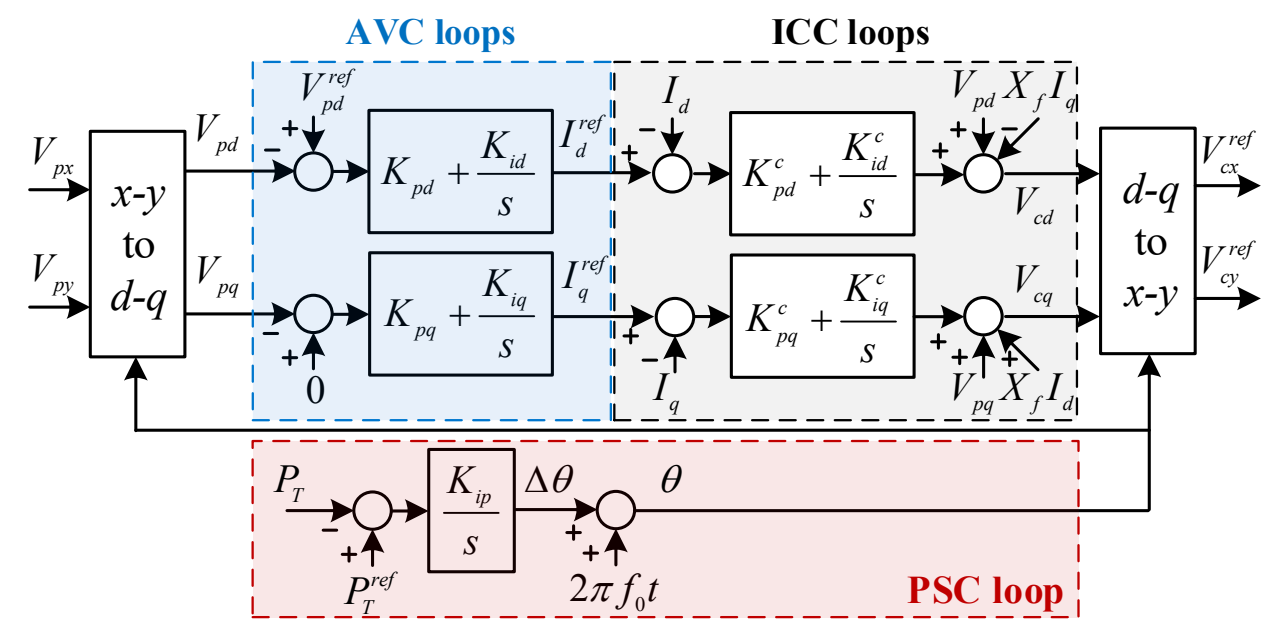

Fig. 6. Configuration of the GFM control.

The relationship between the *d-q* and *x-y* coordinates is illustrated in Fig. 7. In Fig. 7, the *d-q* coordinate is rotating at the angular speed $\omega_p$, highlighted in red. This angular speed is dependent on the PSC loop, denoted as $\theta = \omega_p t$. The *x-y* coordinate, highlighted in blue, is rotating at the angular speed $\omega_v$. The values on the *x* and *y* axes are represented as $V_{px}$ and $V_{py}$, respectively. At steady state, the variable in the *x-y* coordinate, $V_{px} + jV_{py}$, is located on the *d*-axis of the *d-q* coordinate, and the angular speed of the *x-y* coordinate is the same as that of the *d-q* coordinate, i.e., $\omega_p = \omega_v$.

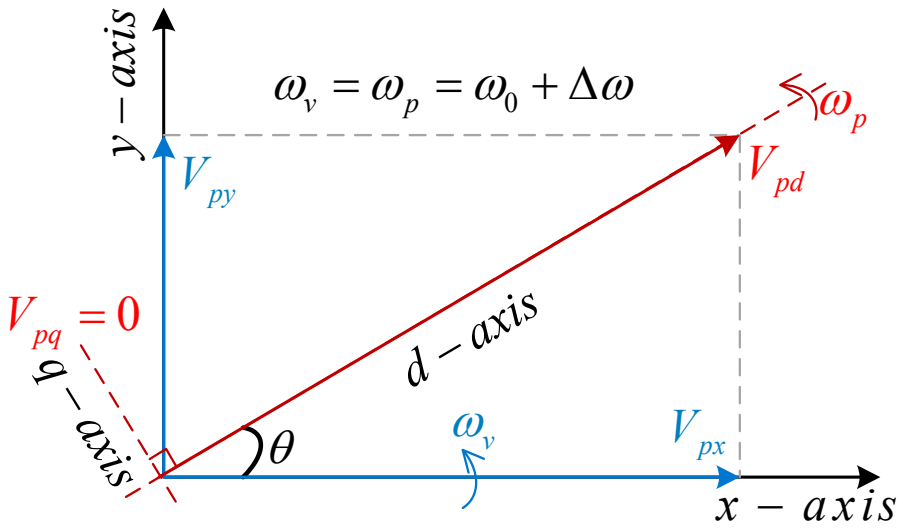

Fig. 7. Relationship between the *d-q* and *x-y* coordinates.

From Fig. 6, transfer function model of the GFM-VSC can be obtained in (1)

$$\Delta \mathbf{V}_{pxy} = \mathbf{G}_{vsc}(s)\Delta \mathbf{I}_{xy} \tag{1}$$

where $\Delta \mathbf{V}_{pxy} = \begin{bmatrix} \Delta V_{px} & \Delta V_{py} \end{bmatrix}^T$, $\Delta \mathbf{I}_{xy} = \begin{bmatrix} \Delta I_x & \Delta I_y \end{bmatrix}^T$,

$$\mathbf{G}_{vsc}(s) = \frac{\mathbf{T}_{xy2dq} + \mathbf{T}_s K_{ip}(s)\mathbf{V}_{pxy0}}{\mathbf{G}_{idq}(s)\mathbf{K}_{vdq}(s)\mathbf{T}_{xy2dq} - \mathbf{T}_s\left(K_{ip}(s)\mathbf{I}_{xy0} + \mathbf{T}_{xy}\right)},$$

$$\mathbf{T}_s = \dot{\mathbf{T}}_{xy2dq}\mathbf{I}_{xy0}{}^T - \mathbf{G}_{idq}(s)\mathbf{K}_{vdq}(s)\dot{\mathbf{T}}_{xy2dq}\mathbf{V}_{pxy0}{}^T,$$

$$\mathbf{I}_{xy0} = \begin{bmatrix} I_{x,0} & I_{y,0} \end{bmatrix}, \quad \mathbf{V}_{pxy0} = \begin{bmatrix} V_{px,0} & V_{py,0} \end{bmatrix}, \quad K_{ip}(s) = -K_{ip}/s,$$

$$\mathbf{T}_{xy} = -\begin{bmatrix} -\frac{V_{py,0}}{\left|V_{p,0}\right|^2} & \frac{V_{px,0}}{\left|V_{p,0}\right|^2} \end{bmatrix},$$

$$\mathbf{K}_{vdq}(s) = -\begin{bmatrix} K_{pd} + K_{id}s^{-1} & \\ & K_{pq} + K_{iq}s^{-1} \end{bmatrix},$$

$$\mathbf{T}_{xy2dq} = \begin{bmatrix} \cos\theta_0 & \sin\theta_0 \\ -\sin\theta_0 & \cos\theta_0 \end{bmatrix}, \quad \dot{\mathbf{T}}_{xy2dq} = \begin{bmatrix} -\sin\theta_0 & \cos\theta_0 \\ -\cos\theta_0 & -\sin\theta_0 \end{bmatrix},$$

$$\mathbf{G}_{idq}(s) = \begin{bmatrix} \frac{K_{pd}^c s + K_{id}^c}{X_f \omega_0^{-1} s^2 + K_{pd}^c s + K_{id}^c} & \\ & \frac{K_{pq}^c s + K_{iq}^c}{X_f \omega_0^{-1} s^2 + K_{pq}^c s + K_{iq}^c} \end{bmatrix}.$$

The dynamics of AC transmission lines are represented by

$$\Delta \mathbf{V}_{pxy} = \mathbf{G}_{line}\Delta \mathbf{I}_{xy} \tag{2}$$

Based on (1) and (2). The oscillation loop is established by the GFM-VSC and AC transmission line. The stability of the SE-GFPS can be assessed according to the solutions of (3):

$$\left|\mathbf{G}_{vsc}(s) - \mathbf{G}_{line}\right| = 0, \text{ or } \left|\mathbf{G}_{vsc}(s)^{-1} - \mathbf{G}_{line}{}^{-1}\right| = 0 \tag{3}$$

where | | means the determinant of the matrix.

## IV. Subsynchronous Damping Analysis of the SE-GFPS

### *A. Self- and Coupling-damping Method*

As the number of considered GFM control loops increases, the investigation of the coupling between these loops and the connected power system becomes more complex. To clarify the causes of SSOs, the self- and coupling-damping method is proposed to identify whether the negative damping originates from the inherent GFM control loops or from the dynamic coupling between the GFM-VSC and the connected power system.

For different GFM configurations, the self-damping remains invariant with respect to the external power system and is solely determined by the GFM control loops, generally being positive with standard design process. In contrast, the coupling-damping varies depending on the scenarios of the connected power systems, as the coupling between the GFM-VSC and the power system is jointly influenced by the control parameters and the operational states. The coupling-damping quantifies the impact of the dynamic interaction between the GFM control loops and the power system and can thus be used to identify the risky factors of oscillations.

In general, the SISO transfer function of GFM control loops can be expressed by selecting an appropriate input and output variable ($\Delta x$ and $\Delta y$, for example) as

$$\begin{gathered} \Delta y = g_0(s)\Delta x \\ g_0(s) = a_n s^n + a_{n-1}s^{n-1} + \ldots + a_2 s^2 + a_1 s + a_0 \end{gathered} \tag{4}$$

where $a_i$ ($i \in [1, n]$) are the coefficients of the transfer function, which depend on the control parameters and operational states of GFM-VSC.

Clearly, there are *n* solutions to (4). In which, denoting the dominant open-loop oscillation mode (the solution with the worst damping) as $\lambda_0 = -\varepsilon_0 + j\omega_0$, the self-damping can be expressed as

$$d_{self} = f_s\left(a_n, a_{n-1}, \ldots, a_2, a_1, a_0\right) \tag{5}$$

Subsequently, considering the dynamics of the connected power system (represented by $\Delta y = g_p(s)\,\Delta x$), the characteristic equation of the power system connected with the GFM-VSC can be derived from (4) as

$$g_0(s) - g_p(s) = \frac{\left(s^2 + d_{self}s + c_0{}^2\right)}{\left(s - \lambda_0\right)\left(s + \lambda_0^*\right)} g_0(s) - g_p(s) = 0 \tag{6}$$

Denoting the dominant closed-loop oscillation mode of (6) (the solution with the worst damping) as $\lambda_s = -\varepsilon_s + j\omega_s$, the coupling damping can be calculated as

$$d_{coup} = -\operatorname{Im}\left[\frac{g_p(\lambda_s)(s - \lambda_0)(s + \lambda_0^*)}{g_0(\lambda_s)}\right] / \omega_s \tag{7}$$

where $\varepsilon_s = d_{sum}/2 = (d_{self} + d_{coup})/2$. $\varepsilon_s$ can be considered zero to correspond to critical stability conditions.

It is evident from (7) that the coupling damping depends on the parameters of both the AC power system and the GFM-VSC, thereby changing the self-damping designed by the GFM-VSC. This explains why oscillations can occur even when the GFM-VSC is inherently stable (i.e., $d_{self} < -d_{coup}$), providing an index for analyzing dynamic interactions between the GFM-VSC and the external power system.

*B. Self-damping of the GFM-VSC*

The self-damping of the GFM-VSC can be determined by analyzing the dynamics of $\mathbf{G}_{vsc}(s)$. To clarify, the form of $\mathbf{G}_{vsc}(s)$ is derived as shown in (8) by establishing the coordinates with reference to the voltage vector of the GFM-VSC, since it acts as a slack bus. Thus,

$$\mathbf{G}_{vsc}(s)^{-1}=\begin{bmatrix} K_d(s) & 0 \\ 2K_{ip}(s)K_q(s)\left(I_{x,0}-K_q(s)\right) & 2K_q(s)-I_{x,0} \end{bmatrix} \tag{8}$$

where $K_d(s)=K_{pd}+K_{id}/s$, $K_q(s)=K_{pq}+K_{iq}/s$.

From (8), the self-damping of GFM-VSC is determined by diagonal elements and is irrelated to off-diagonal elements. This indicates that the PSC loop becomes decoupled from the AVC loops when grid-side dynamic coupling is not considered. The characteristic equation $F_{GFM}(s)$, obtained by $|\mathbf{G}_{vsc}(s)^{-1}|$, is

$$\begin{aligned} F_{GFM}(s) &= K_{pd}\left(2K_{pq}+I_{x,0}\right)s^2+2K_{id}K_{iq} \\ &+\left(K_{id}(2K_{pq}+I_{x,0})+2K_{iq}K_{pd}\right)s=0 \end{aligned} \tag{9}$$

According to (9), the self-damping is defined as

$$d_{self}=K_{id}\left(2K_{pq}+I_{x,0}\right)+2K_{iq}K_{pd} \tag{10}$$

The self-stability holds as long as

$$2K_{pq}+I_{x,0}>0 \tag{11}$$

where all control parameters are positive.

From (11), the self-damping is impacted by the operation state and the AVC loops. Self-damping is well when the GFM-VSC transfers power from the DC network to the SE-GFPS, i.e., when the condition $I_{x,0} > 0$ holds. In cases where the active power flow is from the AC to the DC power system, the self-damping needs concern. To mitigate the instability risk in such scenarios, it is advisable to set the parameter $K_{pq}$ to a value greater than 0.5 p.u., considering that the maximum transmitted current is -1 p.u.

*C. Coupling-damping in the SE-GFPS*

If the grid-side coupling is considered, the characteristic equation is derived as

$$\begin{aligned} &\left|\mathbf{G}_{vsc}(s)^{-1}-\mathbf{G}_{line}^{-1}\right| = F_{GFM}(s)+F_{grid}(s)=0 \\ &F_{grid}(s)=s^2\left[2Y_LK_{ip}(s)K_q(s)\left(I_{x,0}-K_q(s)\right)+Y_L^{\,2}\right] \\ &\mathbf{G}_{line}=\begin{bmatrix} 0 & -X_L \\ X_L & 0 \end{bmatrix},\; Y_L=X_L^{-1} \end{aligned} \tag{12}$$

The oscillation mode of (12) is denoted as $\lambda_s$. Following the damping torque theory [35], the damping contributed to $F_{GFM}(s)$ from $F_{grid}(s)$ is defined as $d_{coup}$, serving as a measure of the effect of dynamic coupling on the oscillation damping. $d_{coup}$ can be obtained by

$$\begin{aligned} d_{coup} &= \operatorname{Im}\left[F_{grid}(\lambda_s)\right]/\omega_s \\ &= -2\varepsilon_s Y_L^{\,2}+2Y_LK_{ip}K_{pq}\left(I_{x,0}+K_{pq}\right)-2Y_LK_{ip}K_{iq}^{\,2}/\left|\lambda_s\right|^2 \end{aligned} \tag{13}$$

From (13), it is evident that the PSC loop is coupled with the AVC loops once GFM-VSC is grid-connected. The coupling strength increases with higher values of the frequency droop coefficient ($K_{ip}$) and the short-circuit ratio (SCR). From (12) and (13), the stability of SE-GFPS is maintained if (14) is satisfied:

$$2K_{pq}+I_{x,0}>0,\; d_{self}>-d_{coup} \tag{14}$$

In (14), the scenario where control parameter $K_{pq}$ is set to a value exceeding 0.5 is considered according to the conclusion drawn from (11).

*D. Damping Sensitivity Analysis with Respect to Inherent and External Impact Factors*

Clearly, the SSO damping is influenced by various factors. Inherent impact factors include the GFM control parameters, whereas external impact factors consist of the SCR and the operating conditions of the SE-GFPS. SSO damping sensitivity analysis is conducted to clarify how these factors collectively affect the closed-loop damping coefficients ($d_{sum} = d_{coup} + d_{self}$).

The impact of $d$ and $q$ axes AVC parameters on the damping can be evaluated by

$$\partial d_{sum}/\partial K_{pd}=2K_{iq},\; \partial d_{sum}/\partial K_{id}=2K_{pq}+I_{x,0} \tag{15}$$

$$\begin{aligned} \partial d_{sum}/\partial K_{pq} &= 2Y_LK_{ip}\left(2K_{pq}+I_{x,0}\right)+2K_{id} \\ \partial d_{sum}/\partial K_{iq} &= -4Y_LK_{ip}K_{iq}/\left|\lambda_s\right|^2+2K_{pd} \end{aligned} \tag{16}$$

From (10) and (15), increasing the $d$-axis AVC parameters enhances the stability by improving $d_{self}$. From (10), (13), and (16), the $q$-axis AVC parameters increase stability by enhancing both $d_{self}$ and $d_{coup}$. Considering the fact that the $d$ and $q$ axes AVC parameters are similar and the reactance of AC lines is small ($Y_L >> 1$), the closed-loop damping is more sensitive to $q$-axis than $d$-axis control parameters.

The impact of PSC and SCR parameters on the damping can be evaluated by

$$\partial d_{sum}/\partial\left(Y_LK_{ip}\right)=2\left(K_{pq}\left(I_{x,0}+K_{pq}\right)-K_{iq}^{\,2}/\left|\lambda_s\right|^2\right) \tag{17}$$

From (13) and (17), the PSC and SCR parameters enhance stability by improving $d_{coup}$. A larger product of them indicates a stronger coupling, which can have more significant effects.

The impact of direction and amplitude of active power on the damping can be evaluated by

$$\partial d_{sum}/\partial I_{x,0}=2Y_LK_{ip}K_{pq}+K_{id} \tag{18}$$

It can be seen from (18) that the power amplitude (related to $I_{x,0}$) directly affects the damping with a linear relationship under constant control parameters and transmission line inductance. This suggests that as the power amplitude increases, the effects become more pronounced. However, its effect on SSO damping is contingent on the power flow direction. Considering the active power is always transmitted from the SEPS to the receiving-end power system, i.e., $I_{x,0} < 0$, a negative effect is introduced to stability when transmitting more active power.

The analysis reveals how these inherent and external impact factors influence the stability of SE-GFPS, that is, by affecting self-damping, coupling-damping, or both. Moreover, specific contributing impact factors are clearly identified. Based on these findings, appropriate oscillation mitigation strategies can be proposed, such as enhancing the self-defense capability or reducing dynamic coupling strength of GFM-VSC.

### E. SSO Mitigation Methods of GFM-VSC

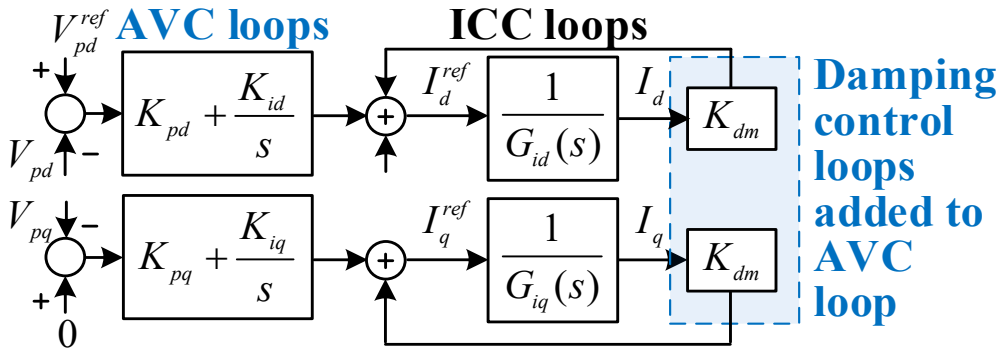

Fig. 8. Configuration of the additional damping control loops.

Considering that the oscillation damping decreases as the output power amplitude increases, an adaptive control for tuning the control parameter corresponding to operational states of the SE-GFPS is proposed, which can be mathematically described as $K_{pq} = K_{pq0} + K_{ad} \times P_T$. However, in practice, similar adaptive controls have not been considered because the tuning of control parameters requires a strict check, which affects the stable operation of the entire SE-GFPS and plays a major role. Therefore, two alternative additional damping control loops are selected as defenses against SSOs. The one is the additional damping control loop added to the PSC loop, the theoretical analysis of which has been conducted well; for interest, refer to references [37]-[38], and we do not repeat here. The other one is the additional damping control loop added to the AVC loops, as illustrated in Fig. 8. This impact of adding these control loops on the SSO damping equals to increase the AVC loop parameters. The justifications are as follows:

From Fig. 8, the transfer function of the AVC and its damping control loops can be written as

$$\begin{aligned}\Delta I_q^{ref} &= -\left(K_{pq} + \frac{K_{iq}}{s}\right)\Delta V_{pq} + K_{dm}\Delta I_q \\ &= -\left(K_{pq} + \frac{K_{iq}}{s}\right)\Delta V_{pq} + \frac{K_{dm}}{G_{iq}(s)}\Delta I_q^{ref}\end{aligned} \tag{19}$$

Therefore, with the additional damping control loops, the equivalent control parameters can be obtained as

$$K_{pq}^{new} = \frac{K_{pq}}{1-K_{dm}}, K_{iq}^{new} = \frac{K_{iq}}{1-K_{dm}} \tag{20}$$

where $G_{iq}(s)$ is considered as 1 considering the fast dynamics of the ICC loops.

From (20), the value of $K_{dm}$ should be in the range of [0,1], and the damping improves as the value of $K_{dm}$ increases.

## V. Case Study

TABLE IV
Parameters of The Case Study

| Parameters | Values |
|---|---|
| Rated AC voltage and Power | 230 kV, 1500 MW (Bipolar) |
| Rated DC voltage | ±500 kV (Bipolar) |
| DC voltage control of VSC-4 | 8 p.u. and 600 rad/s (1.91 p.u.) |
| D-axis AVC loop of VSC-1,2 | $K_{pd}$ = 2.2 p.u. $K_{id}$ = 60 rad/s (0.19 p.u.) |
| Q-axis AVC loop of VSC-1,2 | $K_{pq}$ = 1.4 p.u. $K_{iq}$ = 40 rad/s (0.13 p.u.) |
| PSC loop of VSC-1,2 | $K_{ip}$ = 100 rad/s (0.32 p.u.) |
| Current controls of VSCs | 0.2 p.u. and 100 rad/s (0.32 p.u.) |
| Power modules number per arm | 245 |
| Power module capacitor | 2.58e-04 F |
| AC filter and transmission line | $X_f$ = 0.04 p.u., $X_L$ = 0.075 p.u. |
| DC resistance | 0.0113 Ω/km |
| Simulation and control time | 1e-5 s and 5e-5 s, respectively. |

The configuration of study has been shown in Fig. 5. VSCs 1 and 2 control the AC voltage at 230 kV and frequency at 50 Hz, VSC-4 controls the DC voltage at ±500 kV, and VSC-3 controls the active power at 800 MW. The EMT model is established using the Simulink platform, where the MMC is used and can be accessed by the official example provided in [36]. Some parameters are modified as needed, and which are given in Table IV.

### A. Impact of PSC Parameters on SSOs

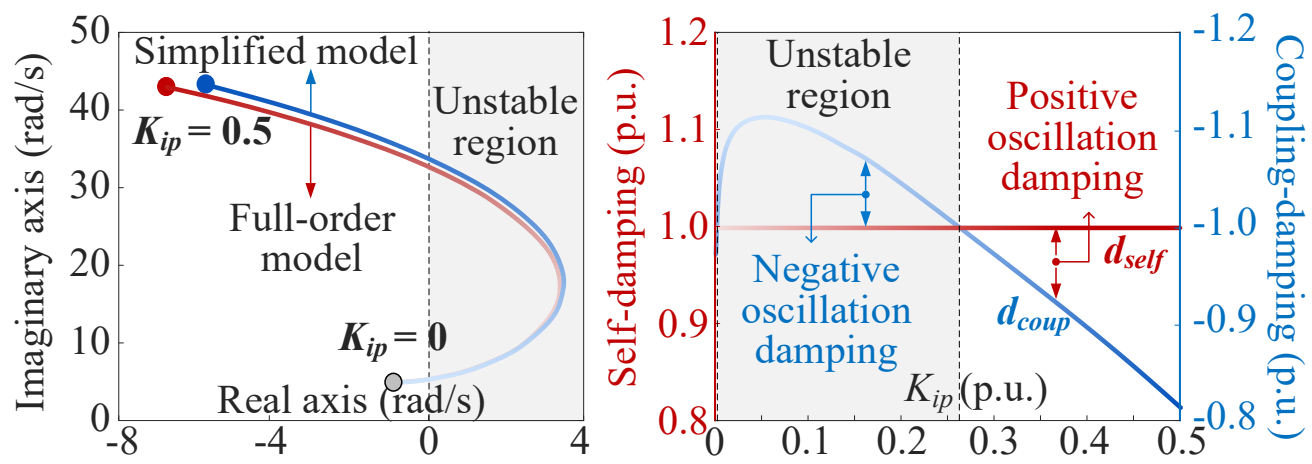

Fig. 9. Trajectories of SSO modes and damping coefficients when $K_{ip}$ varies.

The trajectories of SSO modes while varying $K_{ip}$ from 0 p.u. to 0.5 p.u. are depicted in the left figure of Fig. 9 when the active power is -0.5 p.u. The results from both the full-order model and the simplified model are very close, indicating the accuracy of the simplified model. It shows that oscillation damping initially decreases and then improves as $K_{ip}$ increases, indicating a nonlinear relationship between $K_{ip}$ and oscillation damping. When $K_{ip}$ exceeds 0.26 p.u., stability is restored, and oscillation damping increases with further increases in $K_{ip}$.

The right figure in Fig. 9 demonstrates that adjusting $K_{ip}$ only impacts $d_{coup}$, whereas $d_{self}$ remains constant. As $K_{ip}$ increases, $d_{coup}$ first decreases and then improves. Within the range of 0.002 p.u. to 0.26 p.u. (highlighted in gray area), the negative $d_{coup}$ surpasses the positive $d_{self}$, causing instability. However, when $K_{ip}$ is larger than 0.26 p.u. (as indicated by the black dotted line), the negative $d_{self}$ is eliminated by the positive $d_{self}$, yielding a positive $d_{sum}$ and consequently improving stability.

The EMT simulation results of two cases are given by Fig. 10. During the initial two seconds, $K_{ip}$ is ~ 0 p.u. and the system is stable, as indicated by black curve. However, at 1 s, $K_{ip}$ is changed to 0.18 p.u., observable oscillations occur, as shown by the red curve. And then, at 5.5 s, $K_{ip}$ changes to 0.34 p.u., the system recovers to be stable. These results confirm that the relationship between $K_{ip}$ and oscillation damping is nonlinear, and the mid-range values of $K_{ip}$ should be approached with caution to avoid instability.

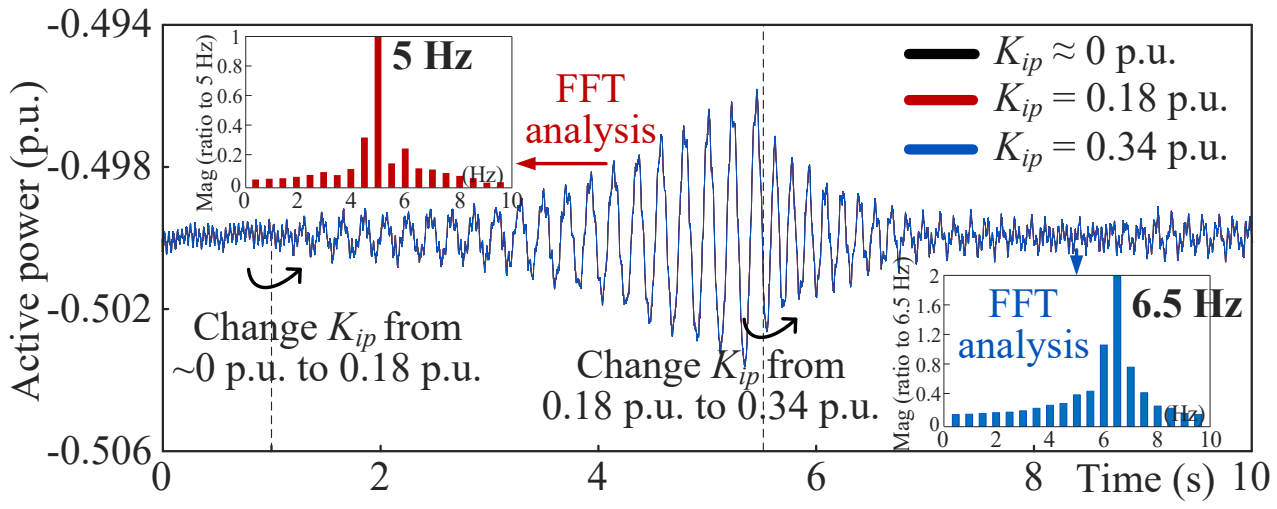

Fig. 10. EMT simulation results when $K_{ip}$ varies.

*B. Impact of AVC Parameters on SSOs*

In this subsection, the $d$- and $q$- axes AVC parameters are involved and the active power remains -0.5 p.u. Firstly, all parameters except for the $d$-axis AVC parameters remain unchanged. The ratio between $K_{pd}$ and $K_{id}$ remains $K_{id} = 0.086 K_{pd}$. The trajectories of SSO modes while varying $K_{pd}$ from 1 p.u. to 5 p.u. are depicted in the left figure of Fig. 11. It is observed that stability gradually improves as $K_{pd}$ increases. The right figure in Fig. 11 demonstrates that increasing $K_{pd}$ enhances $d_{self}$, leading to an increase in oscillation damping. This increase in oscillation damping, in turn, reduces $d_{coup}$ according to (17). Therefore, $K_{pd}$ directly affects $d_{self}$ and indirectly affects $d_{coup}$, and $d_{self}$ is always larger than $d_{coup}$, ensuring the stability.

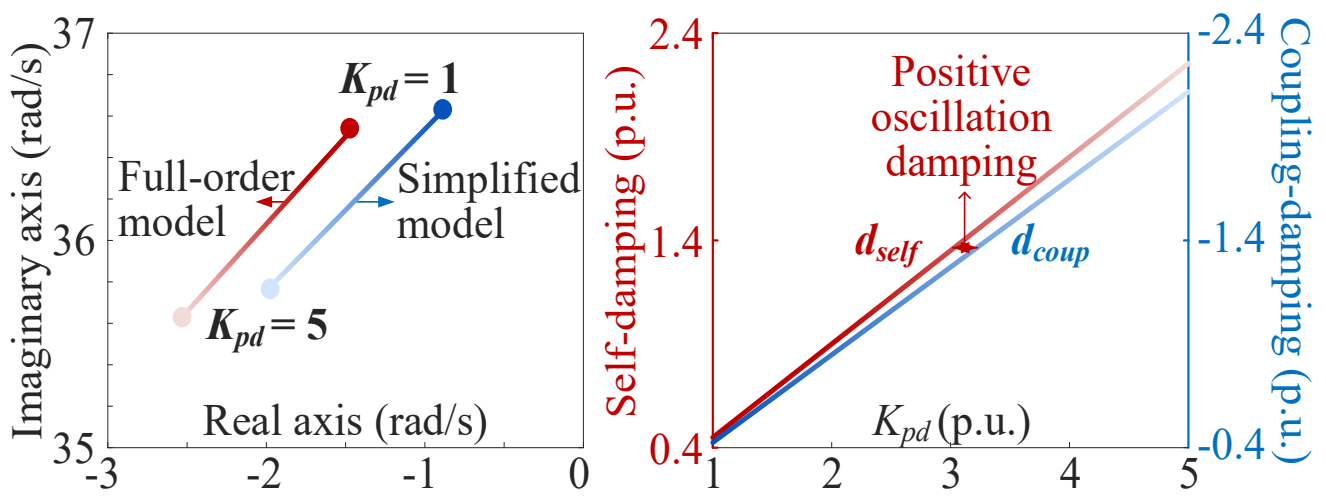

Fig. 11. Trajectories of SSO modes and damping coefficients when $K_{pd}$ varies.

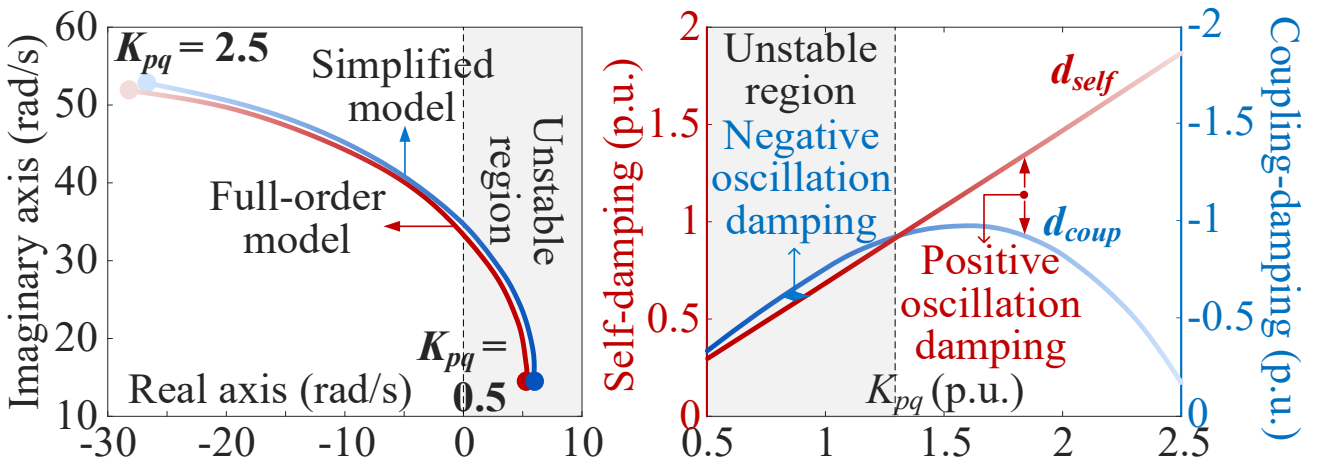

Fig. 12. Trajectories of SSO modes and damping coefficients when $K_{pq}$ varies.

Then, all parameters except for the $q$-axis AVC parameters remain unchanged. The ratio between $K_{pq}$ and $K_{iq}$ remains $K_{iq} = 0.091 K_{pq}$. The trajectories of SSO modes while varying $K_{pq}$ from 0.5 p.u to 2.5 p.u. are depicted in the left figure of Fig. 12. It is observed that changing $K_{pq}$ has a significant impact on oscillation damping, with stability significantly improving as $K_{pq}$ increases. Smaller values of $K_{pq}$ may lead to instability. The right figure in Fig. 12 demonstrates that changing $K_{pq}$ affects stability by altering both $d_{self}$ and $d_{coup}$.

Comparing the results of Fig. 11 to Fig. 12, the influence of $q$-axis control parameters on the SSO damping is markedly different from that of $d$-axis control parameters, as $q$-axis parameters have more pronounced effects on the stability and require greater attention.

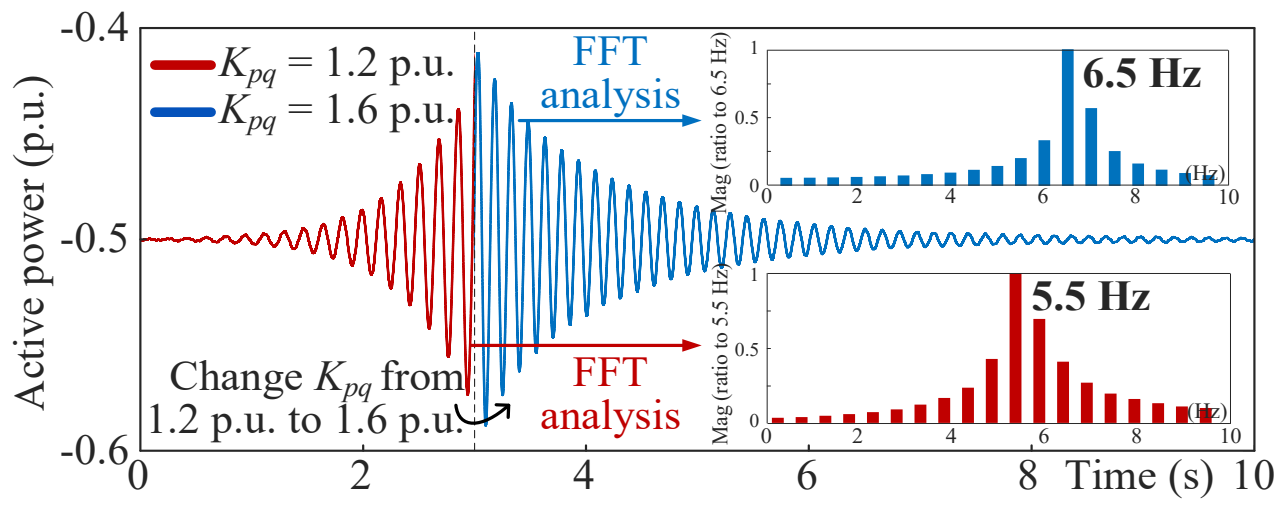

Fig. 13. EMT simulation results when $K_{pq}$ varies.

The EMT simulation results are presented in Fig. 13. The system remains stable with $K_{pq} = 1.4$ p.u. However, at 0 s, $K_{pq}$ is changed to 1.2 p.u., and then noticeable oscillations occur, as illustrated by the red lines. At 7s, $K_{pq}$ is changed to 1.6 p.u., and the oscillation is eliminated. This clearly demonstrates that reducing $K_{pq}$ may increase the risk of instability. It is essential to set $K_{pq}$ to sufficiently high values to ensure an adequate level of damping.

*C. Impact of AC Transmission Line Length on SSOs*

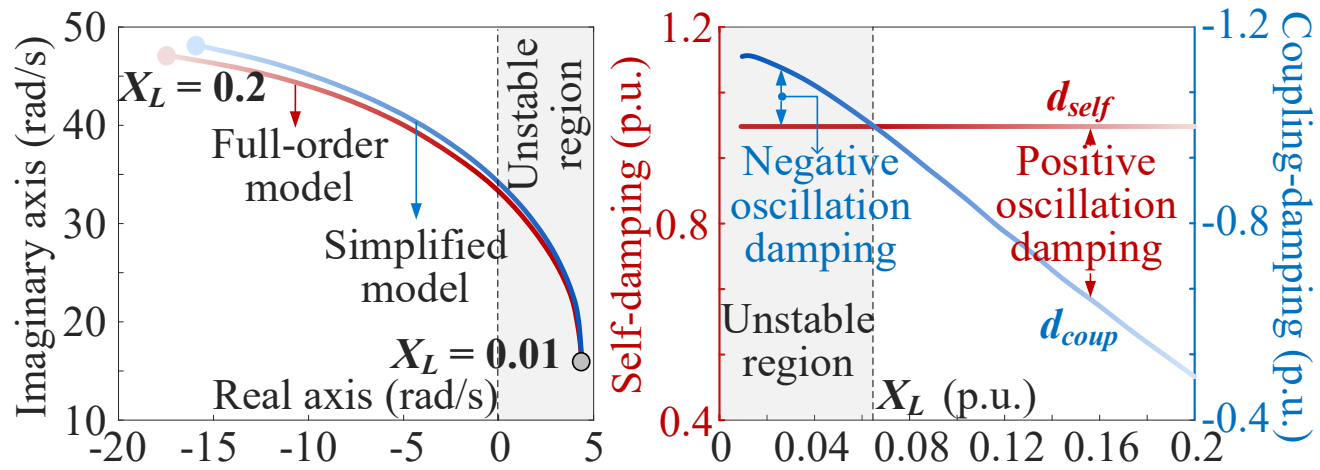

Fig. 14. Trajectories of SSO modes and damping coefficients when $X_L$ varies.

In this case, the parameter $X_L$ is used to analyze the impact of AC transmission line length on the stability. All control parameters remain the same as Table IV, and the active power remains -0.5 p.u. It can be observed from the left figure of Fig. 14 that as $X_L$ reduces (transmission line length shorts), the oscillation damping also decreases. When $X_L$ is smaller than 0.065 p.u., stability is restored. This suggests that the GFM-VSC is better suited for connections with a weaker grid than a significantly stronger grid. Installing a series reactance is beneficial for the stable operation considering the fact that the locations of wind generators are close to the GFM-VSC. The right figure in Fig. 14 demonstrates that changing $X_L$ affects stability by altering $d_{coup}$, whereas $d_{self}$ remains constant. $d_{coup}$ increases with an increase in $X_L$, ultimately resulting in a positive $d_{sum}$ when $X_L$ becomes sufficiently large.

The comparison results of the two cases with different $X_L$ values are illustrated in Fig. 15. The system remains stable with $X_L = 0.08$ p.u. at the initial state. However, at 1 s, $X_L$ is changed to 0.05 p.u., and then noticeable oscillations occur, as illustrated by the red lines. At 3 s, $X_L$ recovers to be 0.08 p.u., and the oscillation is eliminated. These results verify the correctness of conclusions that increasing the reactance for grid connection of GFM-VSC is more beneficial for stability and allows for transmitting more active power under the same conditions.

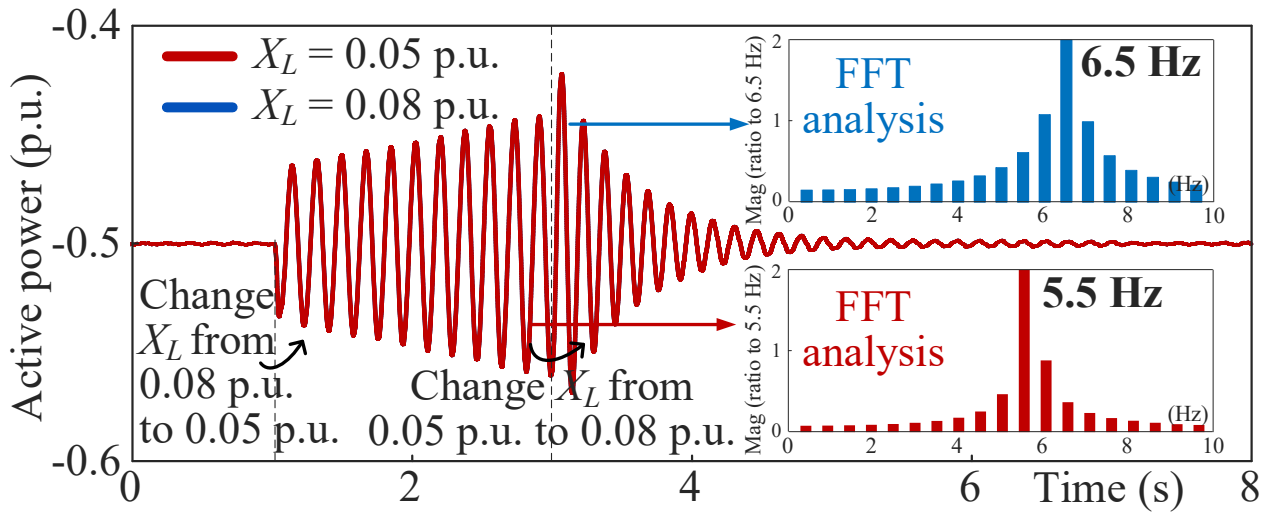

Fig. 15. EMT simulation results when $X_L$ varies.

### D. Stability Enhancement of SE-GFPS

For the proposed adaptive control method, which tunes a control parameter based on the operational states of the SE-GFPS (i.e., $K_{pq} = K_{pq0} + K_{ad} \times P_T$), the parameters are set to $K_{ad}$ = -0.5 p.u., $K_{pq0}$ = 1.4 p.u., and all other parameters match those in Case A. The simulation results are shown in Fig. 16.

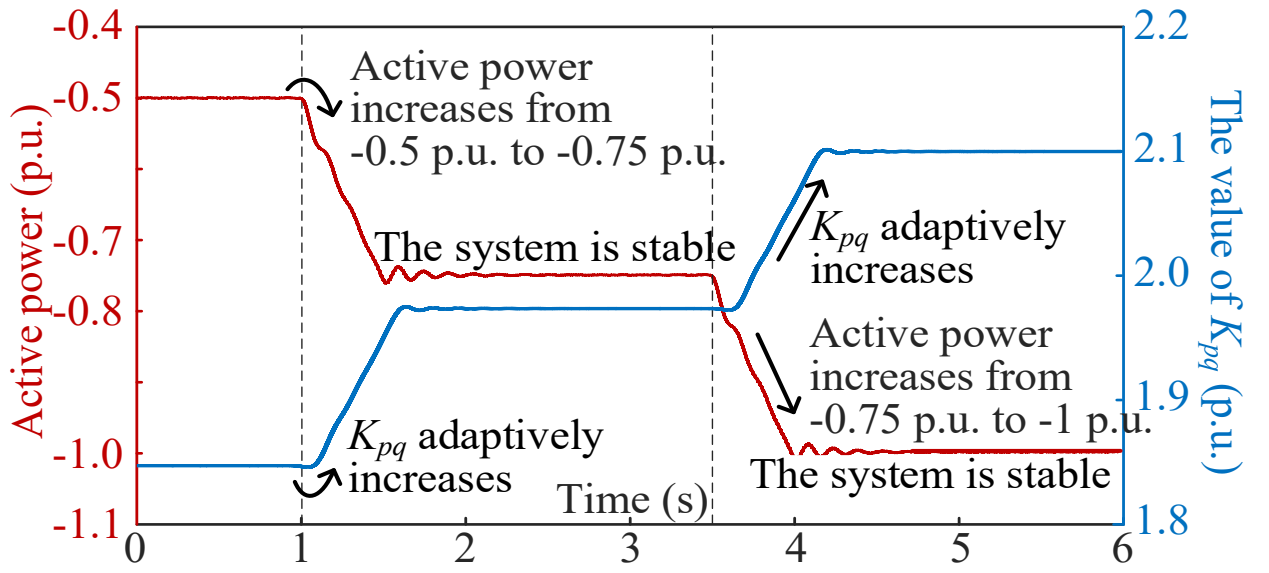

Fig. 16. EMT simulation results with the adaptive control.

In Fig. 16, the curves of $K_{pq}$ and $P_T$ are illustrated. It can be seen that the system is stable at the initial stage, and at 1s, the active power increases from -0.5 p.u. to -0.75 p.u., at the same time, the value of $K_{pq}$ increases as a consequence. At 1.5s, the active power reaches -0.75 p.u., and $K_{pq}$ is 1.975 p.u., the oscillation can be damped well. Then, the active power further increases to -1 p.u., and $K_{pq}$ increases adaptively. At 4 s, the SE-GFPS realizes stable full-power operation, which demonstrates the effectiveness of the proposed parameter-tuning control.

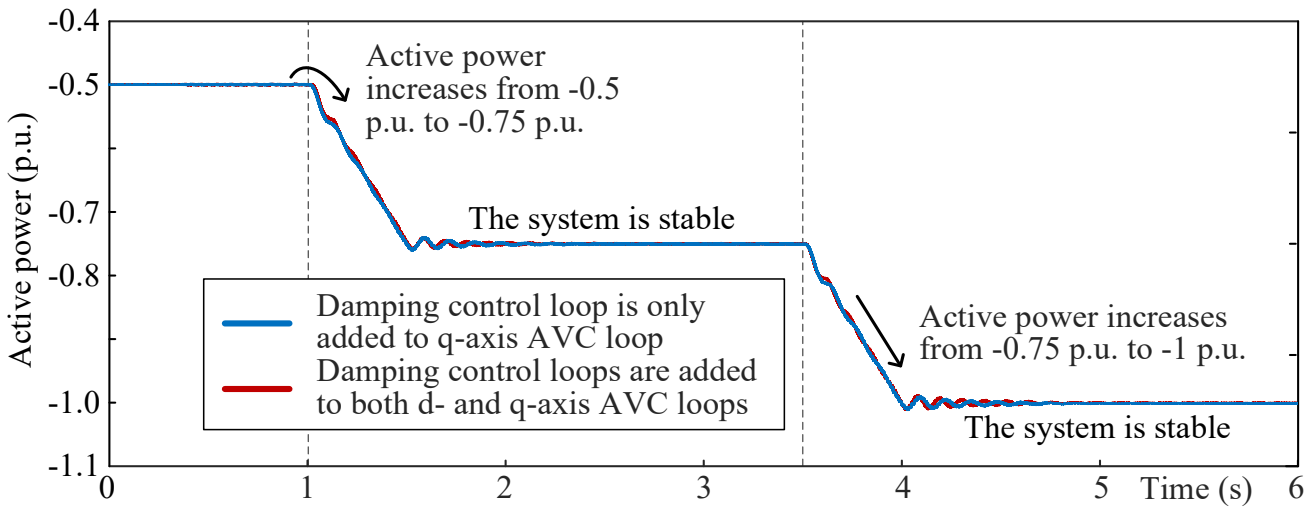

Fig. 17. EMT simulation results with additional damping control loops.

For confirmation of the effectiveness of additional damping control loop added to the AVC loops, the simulation results are illustrated in Fig. 17. It can be seen that the oscillation occurs without the damping control loop, and then, the damping control loop is active, and the oscillation disappears. Although in practice, $d$-axis and $q$-axis are both of damping control loops, with the analysis in this study, only adding the damping control loop to $q$-axis AVC is sufficient since the oscillation is majorly affected by the $q$-axis AVC parameters. For a compassion, the results that only applied to $q$-axis AVC loops are illustrated, demonstrating the correctness of the conclusions, and using the conclusions to further simplify the configuration of the control loops.

## VI. Conclusions

This study addresses the critical issue of SSOs in the SE-GFPS, motivated by practical instances of instability. By investigating the underlying mechanism, the study provides an effective damping analysis method supported by rigorous validation. It offers significant insights into enhancing the stability of SE-GFPS and essential guidance for engineers to identify major risks associated with GFM control in future grid-level applications. The main conclusions are as follows:

1) In SE-GFPSs dominated by GFM-VSC, the GFM control loops play a major role in stability analysis. Therefore, it is important to retain all their dynamics rather than ignore parts of them. This study reports practical examinations that justify the importance of the GFM-VSC in stability analysis. Moreover, the practical GFM control configuration is illustrated, and its full dynamics are retained.

2) This paper proposes a novel damping analysis method by separating oscillation damping into self-damping and coupling-damping. Self-damping is determined solely by the inherent GFM control parameters, while coupling-damping provides a quantitative measure of the dynamic coupling strength between the GFM-VSC and the external power system. These indices offer clear guidance for optimizing the inherent GFM control parameters and selecting appropriate operating conditions for the GFM-VSC.

3) This study clarifies how GFM control parameters affect SSOs. It reveals that the $d$-axis AVC loop parameters primarily affect self-damping, while the PSC loop parameters mainly impact coupling-damping. In contrast, the $q$-axis AVC loop parameters influence both self-damping and coupling-damping. Additionally, the study highlights that DC-side dynamics play a minimal role in the dynamics of SE-GFPS. Therefore, if SSOs are caused by strong dynamic coupling, the parameters of the PSC and $q$-axis AVC loops should be optimized.

4) Through sensitivity analysis, this study identifies critical factors and proposes SSO mitigation methods for stability improvement. It is verified that these methods are equivalent to improving the GFM control parameters. Given that the $q$-axis AVC loop has a more pronounced influence on SSOs compared to the $d$-axis AVC loop, adding damping control to the $q$-axis AVC loop or adaptively adjusting its parameters based on operational variations is sufficient for oscillation mitigation.

These findings provide valuable insights into the stability challenges posed by both current and emerging SE-GFPS, ensuring stable power delivery from isolated renewable energy sources. In future work, the impact of various types of GFM control will be analyzed using the generalized self- and coupling-damping analysis method, which will support the appropriate selection of GFM control strategies for different scenarios.